\documentstyle[editedvolume]{crckapb}

\begin{opening}

\title{A BRIEF HISTORY OF STAR FORMATION AND CHEMICAL ENRICHMENT IN THE BULGE OF THE MILKY WAY}

\author{Jay A. Frogel}
\institute{The Ohio State University\\
            Department of Astronomy\\
            Columbus, OH,  USA}
\end{opening}

\runningtitle{Star Formation in the Bulge of the Milky Way}
\begin{document}

\begin{abstract}
   
Observations of the stellar content of the bulge of the Milky Way can
provide critical guidelines for the interpretation of observations of
distant galaxies, in particular for understanding their stellar content
and evolution.  In this brief overview I will first highlight some
recent work directed towards measuring the history of star formation
and the chemical composition of the central few parsecs of the Galaxy.
These observations point to an episodic history of star formation in
the central region with several bursts having occurred over the past
few 100 Myr (e.g. Blum et al. 1996b).  High resolution spectroscopic
observations by Ram\'{\i}rez et al. (1998) of luminous M stars in this
region yield a near solar value for [Fe/H] from direct measurements of
iron lines.  Then I will present some results from an ongoing program
by my colleagues and myself the objective of which is the delineation
of the star formation and chemical enrichment histories of the central
100 parsecs of the Galaxy, the ``inner bulge''.   From new photometric
data we have concluded that there is a small increase in mean [Fe/H]
from Baade's Window to the Galactic Center and  deduce a near solar
value for stars in the central region.  For radial distances greater
than 1$^{\circ}$ from the Galactic Center we fail to find a measurable population of stars that
are significantly younger than those in Baade's Window.  Within 1$^{\circ}$
we find a number of luminous M giants
that most likely are the result of a star formation episode not more
than one or two Gyr ago.

\end{abstract}
\section{Introduction}
     
     We care about the bulge of the Milky Way both because of
what it tells us about the formation and evolution of our own
galaxy and because its structure and stellar content are often
used as a proxies in the study of other galactic bulges and
elliptical galaxies.  Thus, in the spirit of this meeting, it
serves as a vital link between the near and the far, between the
present and the past.

     Buried within the Galactic bulge is the center of the
Galaxy, a region which, on a small scale, has some properties in
common with luminous AGNs and starburst galaxies.  Because of its
proximity, it can be studied in greater detail than any other
such galactic nucleus.  On the other hand, between us and it lie
clouds of interstellar dust of great enough optical depth that
the average visual extinction is about 30 magnitudes.  Thus it is
only at infrared and longer wavelengths that the central few arc
minutes of the Galaxy can be studied.  Indeed, out to a radius of
about 2$^{\circ}$  (and much farther if observing on or close to
the major axis) the visual extinction is still great enough that
optical observations are difficult to impossible along most lines
of sight.
    
This review of our current state of knowledge of the star formation and
chemical enrichment history of the Galactic bulge will be brief.  I
will make no attempt to cite the many research papers that have
appeared over the past decade or so that deal with these topics.  The
papers that are referred to, though, do contain extensive
references to the literature.  Furthermore, the review will be
restricted to stars and their optical and infrared photospheric
radiation.

Simply put, the history of star formation can be traced by a survey
of either hot blue stars or cool red ones. The former, which will
primarily be massive main sequence stars, are effective tracers of the
most recent epoch of star formation.  The latter will not only be
effective tracers of the most recent epoch -- the late-type supergiants --
but will also trace out older epochs of star formation via the presence
of luminous AGB stars.  This review will deal primarily with surveys of
the cool stellar population in the central part of the Milky Way so
that star formation can be investigated over a broader period of time.
The next section, though, will briefly consider some work on the hot
stellar component in the immediate vicinity of the Galactic Center
itself.

\section{The Central Few Arc Minutes Of The Galaxy}
     
Krabbe et al. (1995) have reported on an extensive survey of the
central few arc seconds of the Galaxy.  They identified more than 20
luminous blue supergiants and Wolf-Rayet stars in a region not more
than a parsec in radius.  The inferred masses of some of these stars
approaches 100 M${\odot}$.  From this they conclude that between 3 and
7 Myr ago there was a burst of star formation in the central region.
They also identified a small population of cool luminous AGB stars from
which one can conclude that there was significant star formation
activity a few 100 Myr ago as well.

Blum et al. (1996a) carried out a K-band survey of the central 2 arc
minutes of the Galaxy and drew renewed attention to the presence of a
significant excess of luminous stars ($K_{0} < 6$) when compared to a
typical old stellar population such as is found in Baade's Window, for
example. Most of these stars were found by Blum and others to be M
stars, presumably a mixture of supergiants and AGB stars.  However, as
Blum et al. (1996a,b) pointed out, the distinction between an M
supergiant and a luminous M-type AGB stars cannot be made on the basis
of luminosity alone since there is a two magnitude range in which the
luminosities of the two very different class of stars overlap (see Fig.
5 of Blum et al. 1996b).  But assigning stars to one class or the other
is critical in deciphering the star formation history of this region.
With K-band spectra, though, it becomes straightforward to make this
distinction for almost all cases (Fig. 1 of Blum et al. 1996b).  As
first quantified by Baldwin et al. (1973) M-type supergiants can be
easily distinguished from ordinary giants of the same temperature (or
color) via the strengths of the H$_{2}$O and CO absorption bands in
K-band spectra.

Blum et al. (1996b) analyzed K-band spectra for a representative
sample of 19 of the luminous stars identified in their survey area.
Only 3 of these stars were found to be supergiants; one of these is the
well known IRS 7.  The remainder are AGB stars, some of which could be
long period variables as well.  From the spectra and the multi-color
photometry they were able to calculate effective temperatures and
bolometric luminosities for the stars.  With the assumption that the
abundance of the stars they observed is comparable to that of disk
stars in the solar neighborhood, they were able to estimate ages for
the stars from a comparison with stellar interior models.  Rather than
continuous star formation, they concluded that there have been multiple
epochs of star formation in the central few parsecs of the Galaxy.  The
most recent epoch, less than 10 Myr ago, corresponds with that found by
Krabbe et al.  (1995). Blum et al. also identified significant periods
of star formation as having occurred about 30 Myr, between 100 and 200
Myr, and more than about 400 Myr in the past.  The majority of stars
they observed are associated with the oldest epoch of star formation.

What about the abundances of stars in the central few
parsecs? Ram\'{\i}rez et al. (1998) have obtained high resolution K
band spectra for 10 M giants in this region and did a full
spectral synthesis analysis of them.  They
were able to measure a true [Fe/H] with their observations of
iron lines and thus remove any ambiguity that could arise by
inferring  [Fe/H] from measurements of elements that are often
used as proxies (e.g. Mg or Ca).  For these 10 stars they derive
a mean [Fe/H] of 0.0 with a dispersion no larger than their
uncertainties, about $\pm 0.2$ dex.  Their mean value is a few tenths
of a dex greater than the mean [Fe/H] determined for Baade's
Window K giants (Sadler et al. 1996; McWilliam \& Rich 1994).
While it may be surprising that the mean [Fe/H] at the Galactic
Center is not super-solar, the small increase in the mean value
of [Fe/H] compared with Baade's Window is consistent with
estimates for the gradient in [Fe/H] in the inner Galactic bulge
(Tiede et al. 1995; Frogel et al. 1999).  On the other hand, a
non-detectable  dispersion in [Fe/H] stands in contrast to a
dispersion that is more than an order of magnitude in size for
the K giants in Baade's Window (Sadler et al. 1996; McWilliam \&
Rich 1994).  It is, however, consistent with the lack of
dispersion found for the M giants in Baade's Window (Frogel \&
Whitford 1987; Terndrup et al. 1991).
   
  The fact that [Fe/H] is near solar at the Galactic Center
with a star formation rate per unit mass at least at present 
is considerably in
excess of the solar neighborhood value suggests that the rate of
chemical enrichment has been quite different at the two
locations.  

The difference in the measured dispersions between K
and M giants remains to be explained. In Baade's Window there is
no detectable population of K giants with luminosities great
enough to place them near the top of the giant branch (DePoy et
al. 1993).  At the same time, it is generally thought that in a
stellar population most of whose stars have [Fe/H] greater than
--1.0, nearly all K giants eventually evolve into M giants.  Thus
the observed dispersions should be similar for the two groups.
That they are not could imply that estimates for evolutionary
rates and lifetimes near the upper end of the RGB and AGB are
wrong.  It could also point to problems with the analysis of the
M giants, although in the case of the Ram\'{\i}rez et al. work this
seems unlikely since the underlying principles of their analysis
are basically the same as that employed for the optical studies
of the K giants.

\section{The Inner Galactic Bulge}
     
Now we turn our attention to the inner $3^{\circ}$ of the
Galactic bulge.  This region, which is interior to  Baade's
Window, will be referred to as the inner Galactic bulge.  With
the 2.5 meter duPont Telescope at Las Campanas Observatory I have
obtained JHK images of 11 fields within the inner bulge, three of
which are within $1^{\circ}$  of the Galactic Center.  The two
questions that are being addressed are:   What is the abundance
of the stellar population in this region and is there any
evidence that a detectable component of the population is
relatively young, i.e. significantly younger than globular
clusters? To answer the question about stellar abundances my
collaborators and I are taking two independent approaches.  The
first is based on the finding of Kuchinski et al. (1995) that the
giant branch of a metal rich globular cluster in a K, JK color
magnitude diagram is linear over 5 magnitudes and has a slope
that is proportional to its optically determined [Fe/H]. Results
from this part of the study, based on the LCO data, will be
summarized here.  The second approach, which is expected to give
a more detailed and precise answer to the abundance question, and
is based on the analysis of K-band spectra obtained at CTIO of
about one dozen M stars in each of 11 fields.  This is a work in
progress.
  
We have used two indicators to test for the presence of
intermediate age stars in the bulge (i.e. an age not more than a
few Gyr as opposed to closer to 10 Gyr).  The first is a
determination of the luminosity of the brightest stars on the
giant branch of each of the fields.  A sign of a relatively young
age would be if there were stars brighter than those found in
Baade's Window.  The second indicator involves a comparison of
the properties of long period variables in the bulge with their
counterparts in Galactic globular clusters.

\subsection{ABUNDANCES IN THE INNER GALACTIC BULGE}
     
The best ``fixed reference point'' in any attempt to determine
abundances within the inner bulge is the determination by
McWilliam \& Rich (1994) of the mean abundance of a sample of K
giants in Baade's Window based on high resolution spectroscopy.
They found a mean [Fe/H] of about --0.2.  A similar result was found by
Sadler et al.(1996) based on spectroscopy of several hundred K
giants in Baade's Window. Furthermore, both of these independent
analyses agreed that the spread in [Fe/H] in Baade's Window was
considerably greater than an order of magnitude and could be as
large as two orders of magnitude.  Observations of Baade's Window
M giants, on the other hand, both in the near IR and of red TiO
bands (e.g. Frogel \& Whitford 1987; Terndrup et al. 1991)
consistently pointed to a greater than solar abundance with no
measurable dispersion.  The independent estimate of [Fe/H] for
the Baade's Window giants based on the near-IR slope method
(Tiede et al. 1995) differed from the previous determinations in
that they found an [Fe/H] close to the value based on the optical
spectra of K giants.

     The near-IR survey of inner bulge fields has yielded color-
magnitude diagrams that, except for the fields with the highest
extinction, reach as faint as the horizontal branch.  Thus, with
data for the entire red giant branch above the level of the HB we
can apply the technique developed by Kuchinski et al. (1995)
which derives an estimate for [Fe/H] based on the slope of the
RGB above the HB.  Although the calibration of this technique is
based on observations of globular clusters, the applicability of
the method to stars in the bulge was demonstrated by Tiede et
al. (1995) in their analysis of stars in Baade's Window.
Although we were able to estimate, statistically, the reddening
to each field, the method itself is reddening independent since
it depends only on a slope measurement.  Based on 7 fields on or
close to the minor axis of the bulge at galactic latitudes
between $+0.1^\circ$ and $-2.8^\circ$ we derive a dependence of
$\langle$[Fe/H]$\rangle$ on latitude for $b$ between $-0.8^\circ$
and $-2.8^\circ$ of $-0.085 \pm 0.033$ dex/degree.    When
combined with the data from Tiede  et al. we find for $-
0.8^\circ \leq b \leq -10.3^\circ$ the slope in
$\langle$[Fe/H]$\rangle$ is $-0.064 \pm 0.012$ dex/degree, somewhat smaller
than the admittedly crude value derived by  Minniti et al. (1995). An
extrapolation to the Galactic Center predicts [Fe/H] $= +0.034
\pm 0.053$ dex, in close agreement with the result of Ram\'{\i}rez et
al. (1998).  Also in agreement with Ram\'{\i}rez et al., we find no
evidence for a dispersion in [Fe/H].  Details of this work are in
Frogel et al. (1999).
   
Analysis of the K-band spectra of the brightest M giants in
each of the fields surveyed is nearing completion; the results
appear to be consistent with those based on the RGB slope method,
namely, an [Fe/H] for Baade's Window M giants close to the
McWilliam \& Rich value but with little or no gradient as one goes
into the central region. Also, little or no dispersion in [Fe/H]
within each field is visible in the spectroscopic data. Further
work on the calibration of these data must be done before
definitive conclusions can be drawn.
   
In summary, several independent lines of evidence point to
an [Fe/H] for stars within a few parsecs of the Galactic Center
of close to solar. The gradient in [Fe/H] between Baade's Window
and the Center is small -- not more than a few tenths of a dex.
Exterior to Baade's Window there is a further small decline in
mean [Fe/H] (e.g. Terndrup et al. 1991, Frogel et al. 1990;
Minniti et al. 1995).  It remains to be seen whether this
gradient arises from a change in the mean [Fe/H] of a single
population or a change in the relative mix of two populations,
one relatively metal rich and identifiable with the bulge, the
other relatively metal poor and more closely associated with the
halo.  Support for the latter interpretation is found in the
survey of TiO band strengths in M giants in outer bulge fields by
Terndrup et al. (1990) for which they found a bimodal
distribution.  McWilliam \& Rich (1994) proposed an explanation
based on selective elemental enhancements as to why earlier
abundance estimates of bulge M giants seemed to consistently
yield [Fe/H] values in excess of solar. What still remains to be
understood is why even recent measurements of the M giant
abundances do not reveal any evidence for an intrinsic dispersion
in [Fe/H] in any given field. Finally, an issue that needs
further investigation is the degree to which the indirect methods used for
getting at [Fe/H] are influenced by selective element enhancements. 

\subsection{STELLAR AGES IN THE INNER GALACTIC BULGE}
     
If a stellar population has an age significantly younger
than 10 Gyr, say not more than a few Gyr, then stars on the AGB
can reach luminosities several magnitudes brighter than they
would in an older population.  After correction for extinction we
noted that our fields closest to the Galactic Center had
significant numbers of bright, red stars.  With the stars in
Baade's Window as a guide we chose a reddening corrected K
magnitude of 8.0 as the limit to the brightest magnitude
obtainable in an old population and counted the number stars in
each surveyed field brighter than this relative to the number in
a predefined, fainter magnitude interval.  We found that at
radial distances greater than $1.3^\circ$ the ratio was constant
with a value equal to that for Baade's Window.  On the hand, for
the fields closer to the center than $1.0^\circ$  this ratio was
significantly larger, implying the presence of a relatively young
population of stars, not more than a couple of Gyr old.  This is
consistent with Blum et al.'s work on the inner few arc minutes
of the bulge.  Details  are in Frogel et al. (1999)
   
The second test we applied to see if there is evidence for a
young population in the Galactic bulge was to compare the
luminosities and periods of bulge long period variables (LPVs) with
those found in globular clusters (Frogel \& Whitelock 1998).  For
LPVs of the same age, those with greater [Fe/H] will have longer
periods.  LPVs with longer periods also have higher mean
luminosities.  In the past, claims have been made for the presence
of a significant intermediate age population of stars in the
bulge based on the finding of some LPVs with periods in excess of
500--600 days.  It is necessary, however, to have a well defined
sample of stars if one is going to draw conclusions based on the
rare occurrence of one type of star.  The M giants in Baade's
Window are just such a well defined sample (e.g. Frogel \&
Whitford 1987).  Frogel \& Whitelock (1998) presented a detailed
comparison of LPVs in the bulge and in metal rich globular
clusters.  They demonstrated that with the exception of a few of
the LPVs in Baade's Window with the longest periods, the
distribution in bolometric magnitudes of the LPVs from the two
populations overlap completely.  Furthermore, because of the
dependence of period and luminosity on [Fe/H] and the fact that
there has been no reliable survey for LPVs in globulars with
[Fe/H] $> -0.25$, while a significant fraction of the giants in
Baade's Window have [Fe/H] $> 0.0$ (McWilliam \& Rich 1994), the
brightest Baade's Window LPVs can be understood as a
result of this higher [Fe/H] compared with globular clusters.
   
Finally, observations with the Infrared Astronomical
Satellite (IRAS) at 12$\mu$m were used to estimate the
integrated flux at this wavelength from the Galactic bulge as a
function of galactic latitude along the minor axis. Galactic disk
emission was removed from the IRAS measurements with the aid of a
simple model. These fluxes were then compared with predictions
for the 12$\mu$m bulge surface brightness based on observations of
complete samples of optically identified M giants in minor axis
bulge fields (Frogel \& Whitford 1987; Frogel et al. 1990).  No
evidence was found for any significant component of 12$\mu$m emission
in the bulge other than that expected from the optically
identified M star sample plus normal, lower luminosity stars.
Since these stars are themselves fully accountable by an old
population, the conclusion from this study was, again, that most of the 
Galactic bulge has no
detectable population of stars younger than those in Baade's
Window, i.e. younger than an age comparable to that of globular clusters.


\begin{thebibliography}{}

\bibitem[]{}    
Baldwin, J. R., Frogel, J. A.,\ \& Persson, S. E.: 1973, {\it ApJ}, {\bf 184},
     427.
\bibitem[]{}    
Blum, R. D., Sellgren, K.,\ \& Depoy, D. L.: 1996a, {\it ApJ}, {\bf 470}, 864
\bibitem[]{}    
Blum, R. D., Sellgren, K.,\ \& Depoy, D. L.: 1996b, {\it AJ}, {\bf 112}, 1988.
\bibitem[]{}    
DePoy, D. L., Terndrup, D. M., Frogel, J. A., Atwood, B.,\ \& Blum,
     R.: 1993, {\it AJ}, {\bf 105}, 2121.
\bibitem[]{}    
Frogel, J. A.: 1998, {\it ApJ}, {\bf 505}, 659.
\bibitem[]{}    
Frogel, J. A., Terndrup, D., Blanco, V. M.,\ \& Whitford, A. E.
   : 1990, {\it ApJ}, {\bf 353}, 494.
\bibitem[]{}    
Frogel, J. A., Tiede, G. P.,\ \& Kuchinski, L. E.: 1999, {\it AJ},
     submitted.
\bibitem[]{}    
Frogel, J. A.,\ \& Whitelock, P. A.: 1998, {\it AJ}, {\bf 116}, 754.
\bibitem[]{}    
Frogel, J. A.,\ \& Whitford, A. E.: 1987, {\it ApJ}, {\bf 320}, 199.
\bibitem[]{}    
Krabbe, A., Genzel, R., et al.: 1995, {\it ApJL}, {\bf 447}, L95.
\bibitem[]{}    
Kuchinski, L., Frogel, J. A., Terndrup, D. M.,\ \& Persson, S. E.
   : 1995, {\it AJ}, {\bf 109}, 1131
\bibitem[]{}    
McWilliam, A.,\ \& Rich, R. M.: 1994, {\it ApJS}, {\bf 91}, 749.
\bibitem[]{}    
Minniti, D., Olszewski, E.,\ \& Rieke, M.: 1995, {\it AJ}, {\bf 110}, 1686.  
\bibitem[]{}    
Ram\'{\i}rez, S., Sellgren, K., Carr, J. S., Balachandran, S., Blum,
     R.,\ \& Terndrup, D.: 1998, in {\it ASP Conf. Ser.}, The Central
     Parsecs, eds. H. Falcke, A. Cotera, W. Duschl,\ \& F.  Melia, 
      in press
\bibitem[]{}    
Sadler, E. M., Rich, R. M.,\ \& Terndrup, D. M.: 1996, {\it AJ}, {\bf 112}, 171.
\bibitem[]{}    
Terndrup, D. M., Frogel, J. A., and Whitford, A. E.: 1990, {\it ApJ},
     {\bf 357}, 453
\bibitem[]{}    
Terndrup, D. M., Frogel, J. A., and Whitford, A. E.: 1991, {\it ApJ},
     {\bf 378}, 742.
\bibitem[]{}    
Tiede, G. P., Frogel, J. A.,\ \& Terndrup, D. M.: 1995, {\it AJ}, {\bf 110},
    2788.
\end{thebibliography}
\end{document}